\documentstyle[11pt,epsfig]{article}
\def\la{\mathrel{\hbox{\rlap{\hbox{\lower4pt\hbox{$\sim$}}}\hbox{$<$}}}}
\def\ga{\mathrel{\hbox{\rlap{\hbox{\lower4pt\hbox{$\sim$}}}\hbox{$>$}}}}
\def\ep{e^{+}/e^{-}}
\def\three{^{3}\mbox{He}}
\def\four{^{4}\mbox{He}}
\def\ratio{^{3}\mbox{He} / ^{4}\mbox{He}} 
\def\trit{^{3}\mbox{T}}
\def\threequarter{\hbox{$\,^3\!/_4\;$}}
\def\fourthird{\hbox{$\,^4\!/_3\;$}}

\begin{document}
\title{New results from AMS cosmic ray measurements}
\author{M. A. Huang\thanks{Current affiliation: Department of Physics, 
	National Taiwan University, Taipei, Taiwan, R.O.C., 
	E\-mail: huangmh@phys.ntu.edu.tw} \\
	Institute of Physics, Academia Sinica, Taipei, 11529, 
	Taiwan, R.O.C. }
\date{\em Proceeding of \\ 
	The First NCTS Workshop on Astroparticle Physics, \\
	Kenting, Taiwan, Dec., 6-9, 2001 \\
	To be published by World Scientific Pub. }
\maketitle
\vspace{-0.5cm}
\begin{abstract}
The Alpha Magnetic Spectrometer (AMS) is a detector designed to search for 
antimatter in the cosmic rays. The physics results from the test flight in 
June 1998 are analyzed and published. This paper reviews the results in the 
five published papers of the AMS collaboration, updates the current 
understanding of two puzzles, albedo $\ep$ and albedo $\three$, and disscusses
the influence of albedo particles.
\end{abstract}


\section{Introduction}
The Alpha Magnetic Spectrometer, AMS, is a space borne charged particle 
detector~\cite{AMS99} that will be installed on the International Space 
Station in 2005 for three years. The primary goals of AMS are to detect 
antimatter and dark matter, as well as to perform precision measurement of 
primary cosmic rays. These goals are closely related to astroparticle physics, 
especially in the matter-antimatter asymmetry, dark matter candidates, and 
atmospheric neutrino. This review begins with a short introduction to the AMS. 
Section 2 summarizes the physics results from the June 1998 shuttle flight. 
Section 3 reviews the updated information about the two puzzles from AMS 
measurements. Section 4 discusses the influence of albedo particles on 
atmospheric neutrino and space science.

In June 1998, a prototype detector, called AMS01, was flown in space shuttle 
Discovery on flight STS-91. The major components of AMS01 iclude a permanent 
magnet, time of flight detectors, and silicone trackers to provide measurements
 of charge, velocity and curvature. Thus it can identify particles and 
antiparticles. In addition, there is an aerogel Cerenkov threshold detector, 
which helps distinguish leptons from hadrons at high energy.

The AMS collaboration is constructing a new detector called AMS02. 
AMS02 is upgraded with the following features.
\begin{list}{$\bullet$}{\itemsep 0cm \parsep 0cm}
\item A super-conducting magnet replaces the permanent magnet. This would 
	increase the magnetic field strength and maximum detectable rigidity 
	by 10 times.
\item A 8 layered silicone tracker replaces the 6 layered one. This would 
	improve rigidity resolution.
\item A ring image Cerenkov counter (RICH) replaces the aerogel Cerenkov 
	counter. 
\item An electromagnetic calorimeter and a transition radiation detector are 
	added.
\item A synchrotron radiation detector is being tested and will be added 
	if it performs well.
\end{list}

\section{AMS physics results}
During the 1998 test flight, AMS01 recorded approximately $10^8$ events. The 
physics results had been published in five journal 
papers~\cite{AMS99,AMS00a,AMS00b,AMS00c,AMS00d}. One recent result of deuteron 
is presented at the 27th International Cosmic Ray Conference~\cite{Lamana}. 
The results related to cosmic rays are reviewed in this section. Detailed 
information about data selection and background elimination can be found in 
original articles or in the review paper~\cite{MAH-TAW7}.

\subsection{Search for anti-helium}
One of the major problems in the evolution of early universe is the 
disappearance of antimatter. The four requirements for matter-antimatter 
asymmetry are not fully complied.  
There are no positive evidences supporting the existence or absence of 
antimatter. A direct detection of anti-nuclei such as anti-helium or 
anti-carbon could signal the existence of antimatter. 

\begin{figure}[htbp]
\begin{center}
\epsfxsize=25pc \epsfbox{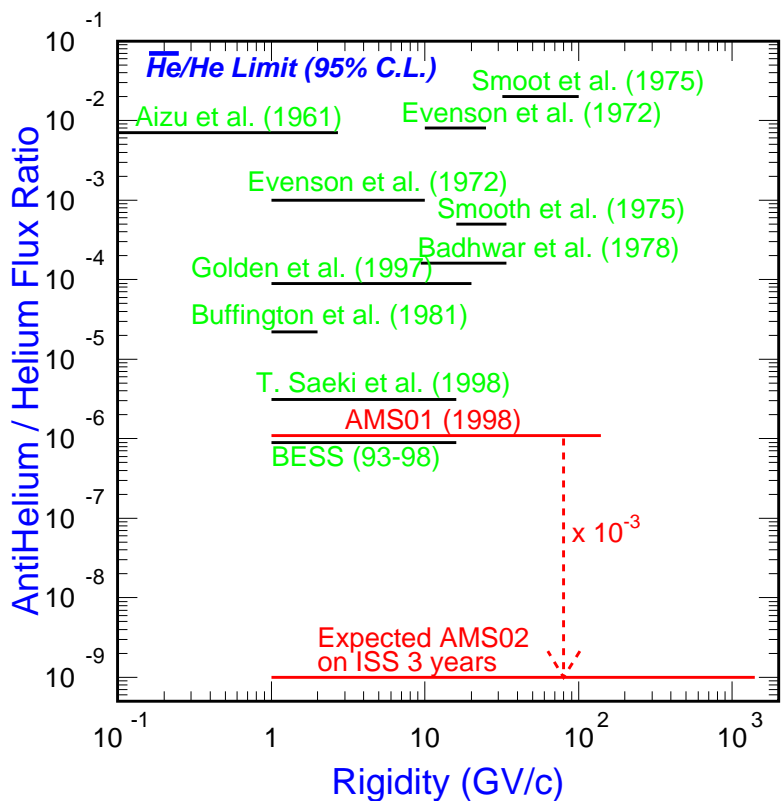} 
\end{center}
\caption{The AMS anti-helium limit is plotted with some previous measurements. 
	These limits assume that anti-helium has the same spectrum as helium.
\label{fig:all-ahe}}
\end{figure}

Under strict selection criteria, no anti-helium was found and $2.86\times 
10^6$ helium with rigidity of 1 to 140GV survives~\cite{AMS99}. The antimatter
 limit at 95\% confidence level is then estimated by assuming that anti-helium 
has the same spectrum as helium. The anti-helium limit is 
$\overline{\mbox{He}}/{\mbox{He}} = 1.1\times 10^{-6}$ 
in the rigidity range of 1 to 140 GV. This result and some previous 
limits~\cite{antimatter} are plotted in Fig.~\ref{fig:all-ahe}. With the 
upgrade in magnet and longer operation time, the AMS02 could reach the 
anti-helium limit to $10^{-9}$, three order of magnitude lower than that of 
AMS01.

\subsection{Search for dark matter}
Recent observation of CMB anisotropy measurements confirmed, once 
again, the existence of a large amount of dark matter. One of the candidates 
of dark matter, weakly interacting massive particle (WIMP), could annihilate 
in the halo of galaxy and produce an excess of positrons. 
The AMS can make indirect search of WIMP through the detection of positrons.

Cosmic positrons come mainly from the decay of charged pions, which are 
produced by interaction of cosmic rays with interstellar medium. Some of the 
early measurements of cosmic positron fraction show a suspicious peak above 
the expected flux of secondary origin at above 10 GeV~\cite{high_e+}. 
However, one recent high statistics experiment~\cite{Barwick95} fail to 
reproduce previous results.

\begin{figure}[htbp]
\begin{center}
\epsfxsize=25pc \epsfbox{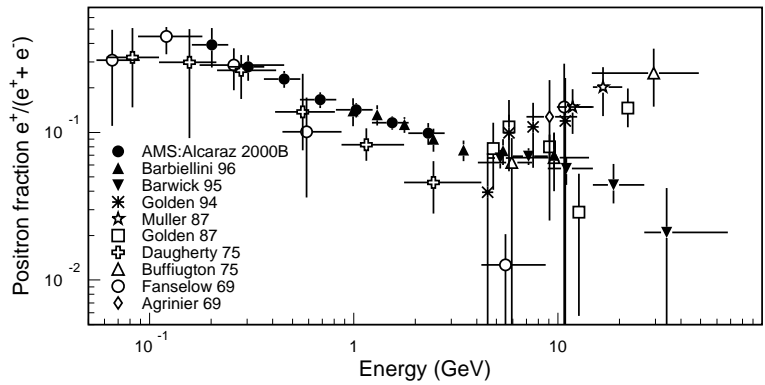}
\caption{The positron fraction of primary cosmic rays measured by the AMS and 
some previous measurements. At energy $<3$ GeV, the AMS and CAPRICE show 
consistent results. \label{fig:positron}}
\end{center}
\end{figure}

For AMS01 data, the separation of positrons from large background of protons 
is limited by the poor performance of the aerogel Cerenkov counter. The energy 
range is only up-to 3 GeV. The AMS fluxes of positron and 
electron~\cite{AMS00b} and the positron fraction, $e^+/(e^+ + e^-)$ are 
consistent with most previous measurements~\cite{Bar96}. 
Fig.~\ref{fig:positron} shows the cosmic positron fraction of AMS01 and several
 previous measurements~\cite{high_e+,Barwick95,Bar96,Daugherty75}. 

The AMS01 cannot identify the possible positron signal from annihilation of 
WIMP at higher energy. The new AMS02 detector will add a ring imaging Cerenkov 
detector and a calorimeter to enhance the chance of detecting this dark 
matter signal.

\subsection{Cosmic rays spectra}

\begin{figure}[htpb]
\begin{center}
\epsfxsize=25pc \epsfbox{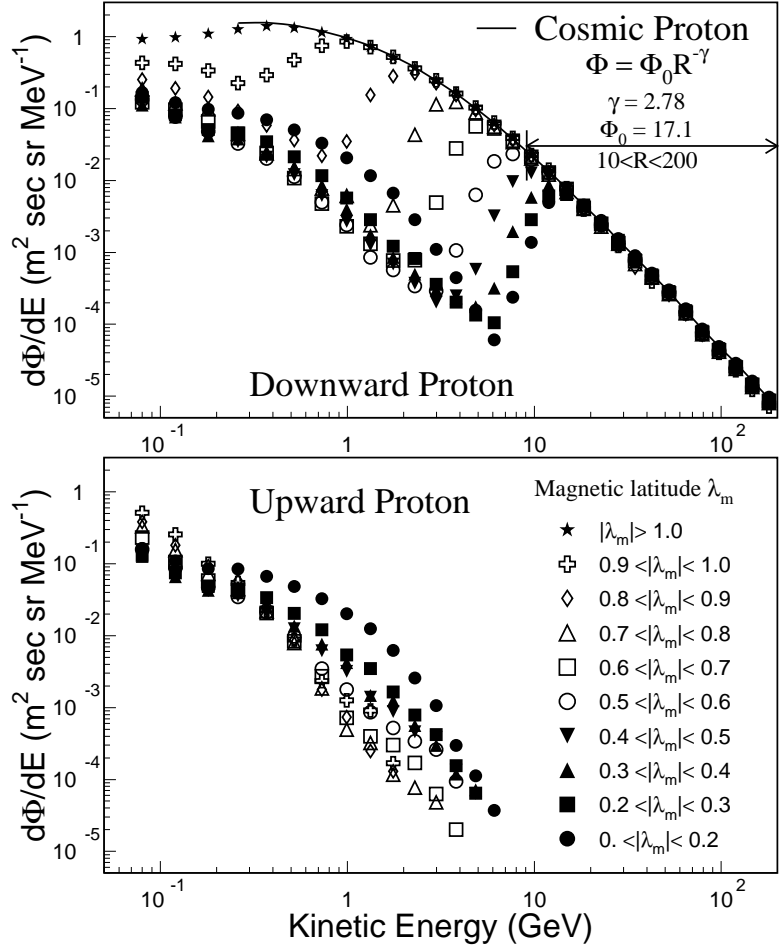}
\caption{The upper figure shows the downward proton fluxes in 10 geomagnetic 
latitude intervals. The solid line is the cosmic proton. For each latitude 
interval, the proton fluxes have a dip due to the rigidity cutoff; below this 
cutoff, there is a second spectrum. The lower figure shows the upward proton 
fluxes. Cosmic rays do not exist in these upward events. 
\label{fig:pflux}}
\end{center}
\end{figure}

Atmospheric neutrinos come from the interaction of cosmic rays with the 
atmosphere. The large acceptance and multiple sub-detectors of AMS can make 
precise measurements of cosmic rays flux and composition. Although the primary 
cosmic ray flux has been measured many times, the AMS is the first instrument 
that measures cosmic rays globally. This information is essential to the 
 calculation of atmospheric neutrino. 

\noindent \underline{\em Proton spectrum}
The first study of protons~\cite{AMS00a} use data from two periods, one with
the detector facing space (downward events) and one with the detector facing 
the Earth (upward events). The data are separated into 10 latitude bins, shown 
in Fig.~\ref{fig:pflux}. For each bin, the spectrum is a mixture of two 
spectra, a cosmic ray and a sub-cutoff component.  
Section 3 will discuss the sub-cutoff components in detail. 

\begin{figure}[htbp]
\begin{center}
\epsfxsize=25pc 
\epsfbox{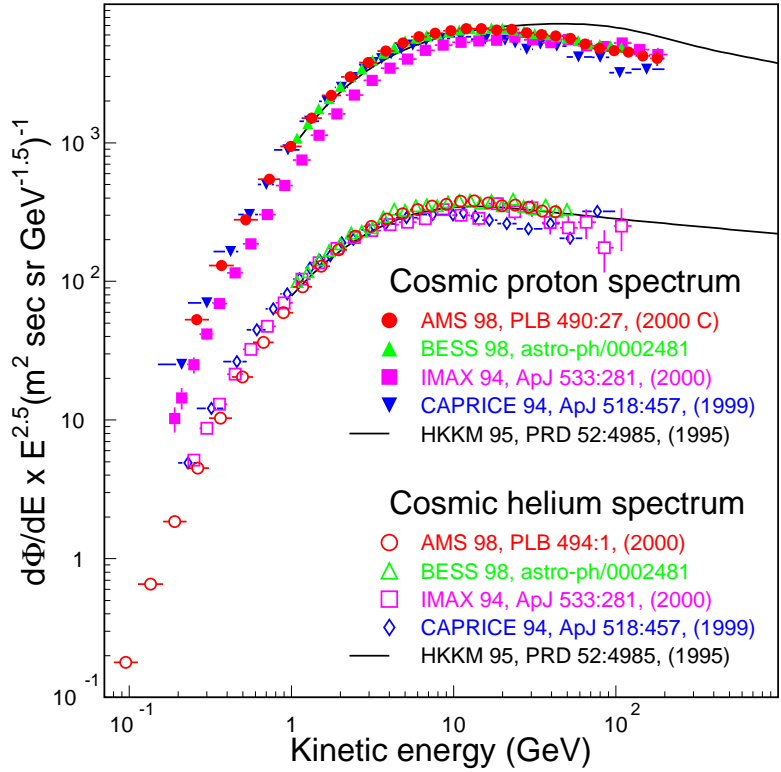}
\caption{The cosmic proton and helium fluxes measured by the AMS are plotted 
	with some previous measurements. The solid lines are the primary 
	proton and helium fluxes used in the HKKM-95 atmospheric neutrino 
	model. \label{fig:p_he}}
\end{center}
\end{figure}

\noindent \underline{\em Cosmic proton spectrum}
All the available data are used in a separate study~\cite{AMS00c} on primary 
cosmic ray proton. The rigidity is selected with
\[ R > (1+2\sigma_{R_c})\times R_{C} \]
where $R_c$ is the maximum of rigidity cutoff in the corresponding geomagnetic 
latitude, and the $\sigma_{R_c}$ is the relative rigidity resolution at $R_c$. 
The final spectrum, shown in Fig.~\ref{fig:p_he}, is fitted to the power law 
spectrum at rigidity $10<R<100$ GV.
\begin{equation}	\frac{d\phi}{dR} = \phi_0 \times  R^{-\gamma}	\end{equation}
The differential spectrum index 
$\gamma$ is $2.78 \pm 0.009 \mbox{(fit)} \pm 0.019 \mbox{(sys)}$ 
and the normalization constant 
$\phi_0$ is $17.1 \pm 0.15 \mbox{(fit)} \pm 1.3 \mbox{(sys)} \pm 1.5 (\gamma)$ 
 $\mbox{GV}^{2.78} /(\mbox{m}^2\; \mbox{s}\; \mbox{sr}\; \mbox{MeV})$.

\noindent \underline{\em Cosmic helium flux}
For the study on cosmic helium flux~\cite{AMS00d}, 79 hours of data taken 
before shuttle landing were used. Helium samples were selected with charged 
number $|Z|=2$. The major contamination comes from protons with 
mis-reconstructed charge, and it is estimated to be less than $10^{-4}$. The 
acceptance of detector was determined to be 0.1 m$^2$ sr at rigidity $>$ 20 GV 
and increased to 0.16 m$^2$ sr at lower rigidity. The overall uncertainty of 
acceptance is 6\%, which include uncertainties from trigger condition (4\%), 
track reconstruction (3\%), and particle interactions combined with event 
selections (2\%). The cosmic ray events are selected when the geomagnetic 
rigidity cutoff of the $\hat z$ of AMS01 detector is less than 12 GV. The 
differential flux, shown in Fig.~\ref{fig:p_he}, was fitted to a power law 
spectrum at rigidity 20 GV to 200 GV.
The differential spectrum index $\gamma$ is 
$2.740 \pm 0.010 \mbox{(stat)} \pm 0.016 \mbox{(sys)}$ and the normalization 
constant $\phi_0$ is 
$2.52 \pm 0.09 \mbox{(stat)} \pm 0.13 \mbox{(sys)} \pm 0.14 (\gamma)$ 
 $\mbox{GV}^{2.74} /(\mbox{m}^2\; \mbox{s} \;\mbox{sr}\; \mbox{MV})$. 

\noindent \underline{\em Influence of AMS cosmic ray measurement}
Fig.~\ref{fig:p_he} shows the cosmic 
proton and helium spectra of AMS, several recent measurements, and spectrum 
used in the atmospheric neutrino 
calculation model. The AMS spectra are consistent with those of previous 
measurements; however, the HKKM-95 model~\cite{HKKM95} seems to have higher 
flux at energy above 20 GeV. Since the cosmic ray flux is the main input 
parameter of atmospheric neutrino simulations, it is difficult to compare the 
difference between results from groups using different models. Inspired by the 
consistency between recent high statistic measurements from AMS~\cite{AMS00c}, 
BESS~\cite{BESS98}, and CAPRICE~\cite{Caprice}, some groups proposed to use an 
unify spectrum~\cite{Gaisser01}. 

\subsection{Cosmic ray light isotopes abundance} 
Cosmic light isotopes such as $^2$D and $\three$ play an important role in 
determining the mean amount of matter traversed by cosmic rays inside the 
galaxy. The excess of $^2$D and $\three$ comes from the spallation of heavy 
cosmic rays, $\four$ or CNO, interacting with interstellar medium (ISM). 
$\four$ loses one nucleon and becomes $\trit$ or $\three$. At AMS observed 
rigidity $R/n\la 1GV/n$, the $\trit$ half life is only 
$\gamma \times 12.26 \la 20$ years, which is much smaller than $10^{7}$ years, 
the typical residence time of cosmic rays. So $\trit$ decays almost completely 
to $\three$. $\four$ also breaks into two $^2$D.  $^2$D and $\four$ have 
the same rigidity for the same energy/nucleon, therefore, they suffer same 
solar modulation effect. The $^2$D/$\four$ is an important indicators for 
studying cosmic ray transportation in galaxy and solar system.

\noindent \underline{\em Cosmic $\three$ } 
From the cosmic helium samples, the helium mass histogram is fitted with two 
components, $\three$ and $\four$. The result shows that 11.5\% of helium is 
$\three$~\cite{AMS00d}. The cosmic ray flux ratio $\ratio$ is approximately 
13\%, much higher than the primordial abundance $\ratio \sim 10^{-4}$. 

\noindent \underline{\em Cosmic $^2$D} 
The deuteron samples are selected with charge +1 and mass compatible with 
that of deuteron~\cite{Lamana}. 
The $1/P$ histograms for several velocity bands for $\beta$ from 0.4 to 0.85 
are fitted with those of proton and deuteron. Approximately 10\% of deuteron 
samples are proton with wrongly reconstructed velocity. After deducting this 
tail, the chance of contamination from residual background of proton in the 
accepted deuteron samples is less than 1\%.

Approximately $10^4$ cosmic deuteron samples are selected from geomagnetic 
latitude $\lambda_m > 0.9$ rad. The deuteron flux is fitted to the solar 
modulation model~\cite{Gleeson}. The best fit of the data is the Local 
Interstellar Space (LIS) spectrum index 2.75 and modulation parameter 
$\phi=650\pm40$ MV, consistent with the solar condition before the solar 
maximum in 20001. The flux ratio $^2$D/$(\three + \four)$ is employed to 
evaluate the effect of cosmic ray transportation effect. The AMS measurement is
 consistent with the prediction of Stephens~\cite{Stephens89}, who used the 
standard leaky box model, and was not  in favor of some non-standard models 
such as re-acceleration theory~\cite{Seo}.

\section{Atmospheric albedo particles \label{sec:albedo}}
\subsection{Particles trapped inside geomagnetic field}
Charged particles having rigidity below geomagnetic cutoff can be trapped 
inside the geomagnetic field. Some have energy low enough that their lowest 
altitudes are well above the atmosphere and they stay trapped for a long time. 
These particles form the radiation belts and have been studied quite thoroughly
 in the early years of space age. 

AMS flew at an altitude approximately 380 km, well below the 
radiation belts. 
Surprisingly, AMS still observed many particles with rigidity below cutoff. 
For all the particles we studied so far, (including p, $^2$D, He, e$^-$, 
e$^+$), all their spectra contain two components, cosmic rays and sub-cutoff 
particles. Unlike the trapped particles in radiation belts, these sub-cutoff 
particles originate from and return to atmosphere in a very short time, less 
than 20 seconds~\cite{MAH-CJP}. They also have some interesting 
features~\cite{MAH-TAW7,MAH-CJP}. Two puzzling phenomena left unanswered in 
the AMS publications~\cite{AMS00b,AMS00d}.

The secondary albedo particles are produced in air shower. The decay chain of 
$\pi \rightarrow \mu \rightarrow e$, produce positrons, electrons, muons, and 
atmospheric neutrinos. The positron electron asymmetry also creates great 
interest among physicists working on atmospheric neutrino simulation. Several 
groups had developed Monte-Carlo simulation to study the production and 
transportation of albedo particles. 

\subsection{Albedo positron electron ratio}
The first puzzle and the most surprising result from AMS01 is the albedo 
positron electron ratio~\cite{AMS00b}. The flux ratio $\ep$ varies with 
magnetic latitude and can be as large as 4 near the magnetic equator. Some 
balloon experiments, operated in high latitude regions, obtained a ratio of 
approximately 1. Most of the radiation belts experiments in the 60s and 70s 
could not distinguish between electrons from positrons. However, the presence 
of positrons in the radiation belts had been reported as early as 
1983~\cite{Just,Galper}. AMS measures at higher energy ($\sim$ GeV), 
almost one order of magnitude higher than that in previous radiation belts 
experiments. The excess of antimatter raises questions concerning their 
origin. At high latitude, these albedo positrons could have rigidity higher 
than the cutoff and be mistaken as cosmic rays. 

\begin{figure}[htbp]
\begin{center}
\epsfxsize=25pc 
\epsfbox{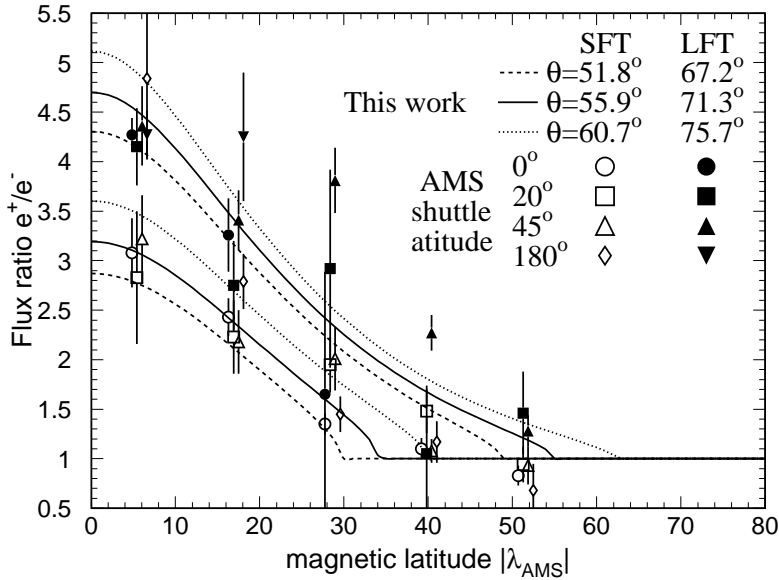}  
\caption{The AMS long flight-time $\ep$ can be explained by the east-west 
effect. The only free parameter, zenith angle $\theta$, in this fitting is 
adjusted according to the atitude of AMS detector. \label{fig:rep}}
\end{center}
\end{figure}

Huang (1998)~\cite{MAH-appc} proposed the first quantitative model of the positron electron 
ratio. For positive charged cosmic rays, the rigidity cutoff
 are lower from the west than that from the east. Therefore, more cosmic rays 
coming from the west than from the east. Because of the geomagnetic field,
 $e^+$ coming from the west and $e^-$ coming from the east have better 
chance to move upward. The combination of higher (lower) fluxes from the west 
(east) and secondary $e^+$  ($e^-$) moving upward produce the positron 
electron asymmetry. The difference in rigidity cutoff decreases with 
increasing magnetic latitude so does the flux ratio.  Using this model and the 
observed $\ep$, Huang (2001)~\cite{MAH-ep} derived the arrival direction of 
primary cosmic rays must be near west (for albedo $e^+$) and east (for albedo
$e^-$), respectively.

Monte-Carlo simulations~\cite{Derome-ep,KMH} had reproduced the $e^+$ and 
$e^-$ fluxes and the $\ep$ as a function of magnetic latitude. They provide
information about the arrival direction of primary cosmic rays. Those 
results confirmed the theoretical prediction from Huang (2001)~\cite{MAH-ep}. 

\subsection{Albedo $\three$}
The second puzzle is that 90\% of albedo Helium is $\three$~\cite{AMS00d}. 
$\three$ had been observed in radiation belts by ONR~\cite{ONR604} at 
kinetic energy 40 - 100 MeV/n and SAMPEX~\cite{Cumming} at kinetic energy 
10-20 MeV/n. Both experiments  observed low energy particles trapped inside 
radiation belts. The AMS observation is in the high energy region and particles
 stay in space for a very short time, $\la$ 20 seconds, compared with the life 
time of trapped particles, which is much longer than days. 

\noindent \underline{\em Spallation of cosmic helium}
Cosmic ray $\four$ nuclei interact with air nuclei, they break up into 
$\three$, whose rigidity would be \threequarter times that of $\four$. When 
the incoming $\four$ has rigidity less than about \fourthird times the cut-off 
rigidity, the $\three$ fragment, having rigidity smaller than the cut-off, 
turns into an albedo~\cite{Lipari}. 

Huang and Stephens~\cite{MAH-he3} simulate the interaction of cosmic helium 
with atmosphere. The result shows that $\three$ produced by spallation exists 
only in specific phase space of rigidity and magnetic latitude.
Only few events near $-0.6 < \lambda_m < -0.5$ and energy $>$1 GeV/n can be 
explained by this mechanism.

\noindent \underline{\em Pick-up Reaction: p ( $\four$, d)$\three$}
Selesnick and Mewalt~\cite{Selesnick} proposed that protons in radiation belts 
picking up one neutron from helium in the upper atmosphere may be 
able to explain the light isotopes in radiation belts. However, the radiation 
belt protons are not energetic enough to produce the $\three$ as observed by 
AMS. Huang and Stephens~\cite{MAH-he3} modified this model using 
cosmic ray protons. Although $\three$ could be produced in low latitude and 
rigidity ranges similar to the AMS measurements, there are some serious 
difficulties. First, the ambient helium density is too low to produce $\three$ 
flux comparable with that measured by the AMS. Second, the spectrum is too 
steep at energy higher than 0.5 GeV/n.

\noindent \underline{\em Monte-Carlo simulation of light isotopes} 
Derome and Bu\'enerd~\cite{Derome-he3} also analyzed the light isotopes, such 
as deuteron, $\three$, $\trit$, from their simulations. The coalescence model 
is used to explain the production of $\three$. The incident proton gets 
absorbed in the atmospheric nitrogen or oxygen to form a compound nucleus, 
which decays to various light nuclei. The individual nucleons must be close to 
each other in order to form a nuclei. Therefore, the probability of forming 
heavier nuclei is much less then that of forming lighter nuclei. 
This model explains successfully the existence of albedo proton, $^2$D, 
$\three$, and $\four$. 
Pugacheva et al.~\cite{Pugacheva} also had a similar explaination of hydrogen 
and helium isotopes in radiation belts. Their model can be employed to 
understand the $\three$ observed by AMS.

\section{Influence of albedo particles}
\subsection{Influence on atmospheric neutrino simulation}
The albedo proton, as shown in the bottom of Fig. 3, could also produce air 
shower and contribute to the atmospheric neutrino flux. Since the albedo flux 
is approximately 1\% of the primary cosmic ray flux, the contribution should be
 in the same order of magnitude or less~\cite{KMH,Lipari}. A good simulation 
should be able to explain all the features of albedo particles, including 
particle types, spectra, spatial and temporal distributions. 

However, Plyaskin~\cite{Plyaskin} claimed that his simulation reproduced the 
albedo positron electron ratio and up-down asymmetry of atmospheric neutrino 
without neutrino oscillation. This simulation uses an approach quite different 
 from the others. Plyaskin used the GEANT simulation code and reduced the size 
of the Earth and the atmosphere into something similar to the size of an 
detector in accelerator experiment. He reported a large excess of contribution 
coming from the scattered protons which enter the atmosphere from the forbidden
 cone, where belong to trapped particles. These scattered particles are not 
counted in traditional simulation algorithms. It is suspicious that the large 
contribution from scattered protons may be the cause of his condensation of 
atmosphere and amy disappear in realistic case.

Although the current simulations reproduce most of the features of albedo 
particles. There remain some questions. The main argument is whether AMS or 
Monte-Carlo simulations over-counted the long flight time (LFT) events. While
the LFT events passed through the AMS altitude many times, they 
were detected in one position only. Some simulations counted the flux as from 
one single particle, others simulations counted each passage through AMS 
altitude. The drift period of albedo particle in GeV range is approximately 10 
second, much shorter than the 90-minute period of the space shuttle. AMS could 
not detect the same particle twice, therefore, AMS did not over-count the LFT 
particles. 

Another question is the precision of simulation. 
The differences between MC simulation and experimental data are much larger 
than the error bars. Discrepency is the greatest near the cutoff and high 
latitude regions. It simply shows that much remains to be learnt concering the 
penumbra region (an intermittent transition zone from trapped particles to 
cosmic rays)~\cite{MAH-TAW7}. There is certainly much room for improvement in 
the Monte-Carlo simulations.

\subsection{Influence to space science}
Space engineers must design proper shielding for satellites or astronauts to 
reduce the ionization radiation caused by the charged particles in space. 
There were models of proton and electron fluxes in radiation belts,
such as AP8 and AE8~\cite{NSSDC}. The albedo particles have energy close to 
the minimum of ionization energy loss and could penetrate deep inside the 
protection layer. This effect had not been considered yet. In recent years, 
space physicists have regained interests in the high energy components in 
radiation belts. With the large acceptance and particle identification 
capability, AMS could be the most powerful detector compared with other 
radiation belt experiments. 

The AMS provide a global measurement of high energy particles, the 
measurements could be used to reconstruct a useful model. However, the current 
data are not very helpful. Although the energy loss in detector materials had 
been restored to the ``original energy'' by deconvolution using Bayes 
theorem~\cite{AMS00a}, this is only a statistical correction. 
Particles entering the detector suffer different amount of energy loss. Also 
the effect of residual magnetic field is not corrected. Those two corrections 
change the pitch angle and energy distributions. So the current flux of albedo 
particles could have large systematic error and can only be used for crude 
estimation. Some of the mismatches in AMS measurements and MC simulations 
might come from the simplification of deconvolution. 

\section*{Acknowledgments}
The author wishes to thank the organizer of this workshop, Professor G.L. Lin 
for his kind invitation. The author was supported by the topical program 
``Detecting cosmic rays with a precise space spectrometer'' from Academia 
Sinica, Taiwan, R.O.C..

\end{document}